\documentclass{article}
\pdfoutput=1

\usepackage{PRIMEarxiv}

\usepackage[utf8]{inputenc} 
\usepackage[T1]{fontenc}    
\usepackage{hyperref}       
\usepackage{url}            
\usepackage{booktabs}       
\usepackage{amsfonts}       
\usepackage{nicefrac}       
\usepackage{microtype}      
\usepackage{lipsum}
\usepackage{fancyhdr}       
\usepackage{graphicx}       
\graphicspath{{media/}}     
\usepackage{float}
\usepackage[numbers]{natbib}
\usepackage{subcaption}

\usepackage{fancyhdr}
\title{Low-latency Real-time Voice Conversion on CPU}

\author{
  Konstantine Sadov \\
  Koe AI \\
  \texttt{ksadov@koe.ai} \\
  \And
  Matthew Hutter \\
  Koe AI \\
  \texttt{mhutter2@washcoll.edu} \\
  \And
  Asara Near \\
  Koe AI \\
  \texttt{asara@koe.ai} \\
}

\begin{document}
\maketitle

\begin{abstract}
We adapt the architectures of previous audio manipulation and generation neural networks to the task of real-time any-to-one voice conversion. Our resulting model, LLVC (\textbf{L}ow-latency \textbf{L}ow-resource \textbf{V}oice \textbf{C}onversion), has a latency of under 20ms at a bitrate of 16kHz and runs nearly 2.8x faster than real-time on a consumer CPU. LLVC uses both a generative adversarial architecture as well as knowledge distillation in order to attain this performance. To our knowledge LLVC achieves both the lowest resource usage as well as the lowest latency of any open-source voice conversion model. We provide open-source samples, code, and pretrained model weights at \url{https://github.com/KoeAI/LLVC}.
\end{abstract}

\keywords{voice conversion \and streaming \and low-latency \and model distillation \and open-source}

\section{Introduction}
Voice conversion is the task of rendering speech in the style of another speaker while preserving the words and intonation of the original speech\cite{walczyna2023}. "Any-to-one" voice conversion converts speech from an arbitrary input speaker which may not have been seen during training to speech in the style of a single fixed speaker. Practical applications of voice conversion include speech synthesis, voice anonymization, and the alteration of one's vocal identity for personal, creative, or professional purposes. 

The core challenges of voice conversion are ensuring similarity to the target speaker and creating natural-sounding output. Real-time voice conversion presents additional challenges that existing high-quality speech synthesis networks are ill-suited for: not only must the network operate faster than real time, but it also must operate with low latency and with minimal access to future audio context. Lastly, real-time voice conversion networks intended for widespread consumer usage must also be able to operate in low-resource computational environments.

This paper proposes an any-to-one voice voice conversion model based on the Waveformer architecture\cite{veluri2023realtime}. While Waveformer is designed to perform real-time sound extraction, LLVC is trained on an artificial parallel dataset of speech from various speakers which have all been converted to sound like a single target speaker with the objective of minimizing perceptible difference between the model output and the synthetic target speech. LLVC is presented as the first open-source model which can convert voices in a streaming manner on consumer CPUs with a latency as low as 20ms.

\section{Related work}
\subsection{Voice conversion}
Early approaches to voice conversion used Gaussian mixture models\cite{shum2008}, with more recent approaches using artificial neural networks\cite{huang2021far} and contemporary architectures commonly including variational autoencoders (VAEs) and generative adversarial networks (GANs)\cite{Qian_2020}. Recent approaches are generally made to operate on non-parallel datasets, referring to datasets where the speakers are not required to perform identical utterances. This is often achieved by a type of bottleneck in the architecture, such as the bottleneck in a VAE\cite{qian2019autovc}, adaptive instance normalization\cite{chen2020againvc}, k-nearest neighbors\cite{baas2023voice} or with the inclusion of pre-trained models which separate content and style, such as automatic speech recognition (ASR) or phonetic posteriorgrams (PPGs)\cite{Liu_2021}.

\subsection{Real-time voice conversion}
There exists several published voice conversion architectures capable of operating at high enough speed to make real-time conversion on consumer hardware feasible. MMVC\footnote{\url{https://github.com/isletennos/MMVC\_Trainer}}, so-vits-svc\footnote{\url{https://github.com/svc-develop-team/so-vits-svc}}, DDSP-SVC \footnote{\url{https://github.com/yxlllc/DDSP-SVC}} and RVC\footnote{\url{https://github.com/RVC-Project/Retrieval-based-Voice-Conversion-WebUI}} are incorporated into the popular real-time voice-changer\footnote{\url{https://github.com/w-okada/voice-changer}} application repository on Github. 

Despite their inclusion in an application dedicated to real-time voice conversion, none of the cited architectures are trained to operate on low-latency streaming audio. Naively converting short sequential segments of audio results in perceptually-degraded output, so the networks are instead adapted for the streaming task by prefixing new input with previous audio context, trading computational efficiency for increased conversion quality.

QuickVC\cite{guo2023quickvc} is capable of running efficiently on CPU and can be adapted to real-time conversion using the same process as the architectures above. Regardless, the absence of streaming-specific architecture leaves this model subject to the same quality and efficiency trade-off as the previously cited models.

The above models share an encoder-decoder structure inspired by VITS\cite{kim2021conditional}. The encoder is comprised of a pre-trained encoder, usually contentvec\cite{qian2022contentvec} or hubert-soft\cite{van_Niekerk_2022}, which are designed to encode speech content without encoding input speaker characteristics such as pitch and timbre. The decoders of MMVC, so-vits-svc, DDSP-SVC and RVC are based on the architecture of HiFi-GAN, while QuickVC uses a vocoder based on the inverse short-time Fourier transform operation\cite{kawamura2023lightweight}.

\subsection{Streaming audio processing}
Neural audio codecs such as LPCNet\cite{valin2019lpcnet}, and EnCodec\cite{défossez2022high} are designed to operate in low-resource streaming settings and have a similar encoder-decoder structure to the real-time voice conversion systems described above. However, these audio codec encoders seek to preserve input speaker identity along with speech content in order to ensure the fidelity of reconstructed audio, and are thus unsuitable for the task of voice conversion.

Waveformer's encoder-decoder architecture is designed to modify input audio by constructing a mask which is added to the input audio signal in order to isolate a type of sound present in the training set, i.e acoustic guitar, coughing, gunshot. While the encoder's initial convolution provides access to a small amount context, dilated causal convolutions (DCC) in the encoder and a masked transformer that attends only to present and past tokens in the decoder ensure that the model's inference is based mostly on past data. This makes the architecture well-adapted for a streaming setting, where requiring future context introduces additional latency. Additionally, the causal nature of the encoder and decoder allow intermediate calculations to be cached for future inference passes, which gives the network access to past context without requiring the entire context to be run through every part of the network, increasing inference speed.

\subsection{Knowledge distillation}
Model distillation in the realm of deep learning refers to the process of utilizing a larger, more complex "teacher" model to supplement the training of a smaller "student" model \cite{DBLP:journals/corr/abs-2006-05525}. This methodology is rooted in harnessing the predictive power of intricate neural architectures while ensuring computational efficiency, especially in scenarios where computational resources are scarce, or when real-time responses are imperative, such as on mobile or edge devices \cite{Alkhulaifi_Alsahli_Ahmad_2021}. Model distillation has recently been utilized to great effect with imitation-trained language models, which use high-quality output of large proprietary models to perform instruction-tuning on smaller open-source language models\cite{taori2023stanford}.

The conventional distillation process involves a teacher model, typically characterized by its large size or loose training constraints, which is trained to perform a specific task using a given dataset, followed by training a student model to mimic the teacher's output distribution, often softened by a higher temperature in the softmax function to encapsulate more nuanced information beyond hard labels \cite{DBLP:journals/corr/abs-2006-05525}. 

In the scenario of non-parallel data, the landscape of model distillation extends to an innovative paradigm. A teacher model is trained on non-parallel data, leveraging its large, complex architecture to sift through and assimilate representations from the inherently unstructured and unaligned data. Following this, a synthetic parallel dataset is engineered based on the teacher's acquired knowledge, which in turn serves as the training ground for the student model \cite{Nagano2021KnowledgeDL, Zhu_Li_Gu_Xu_2023}. 

Although parallel speaker datasets have historically been challenging to create, introducing additional challenges such as aligning the utterances in time\cite{Helander2008OnTI}, the quality of modern voice conversion networks is now high enough such that they can now be artificially created. This can be done by using a pre-existing any-to-one or any-to-many voice conversion network to generate time-aligned parallel voice datasets. These artificial datasets can scale to arbitrary size simply by increasing the amount of input and output pairs generated from inference. After a parallel dataset has been obtained, smaller models can be trained on this dataset which require fewer parameters and less architectural complexity.

\section{LLVC}
\label{sec:methods}

\subsection{Architecture}
Our proposed model is composed of a generator and a discriminator. Only the generator is used at inference time.

\begin{figure}[ht]
\begin{subfigure}[h]{0.3\textwidth}
\centering
\includegraphics[scale=1]{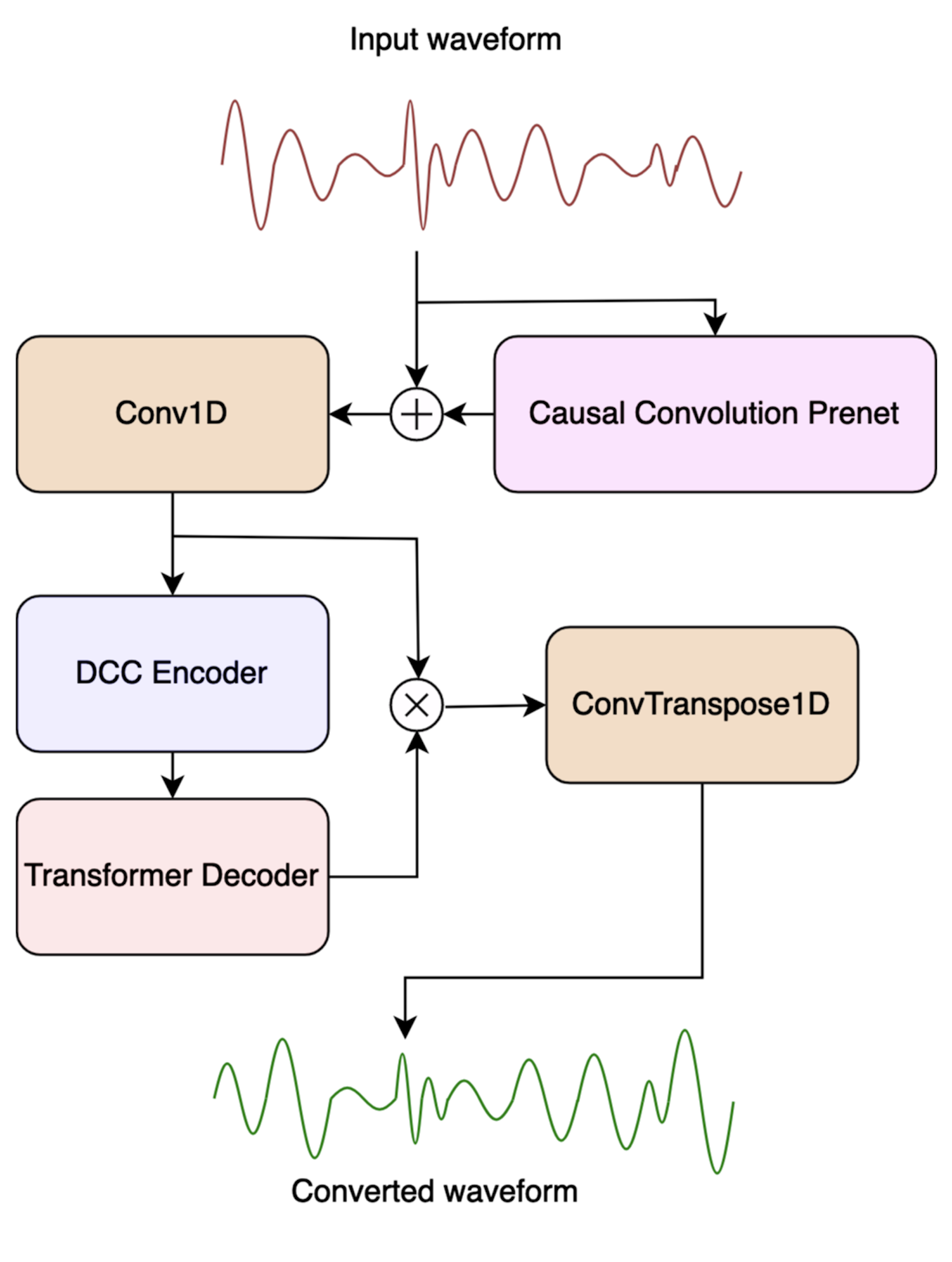}
\caption{Generator}
\end{subfigure}
\hfil
\begin{subfigure}[h]{0.3\textwidth}
\raggedleft
\includegraphics[scale=1]{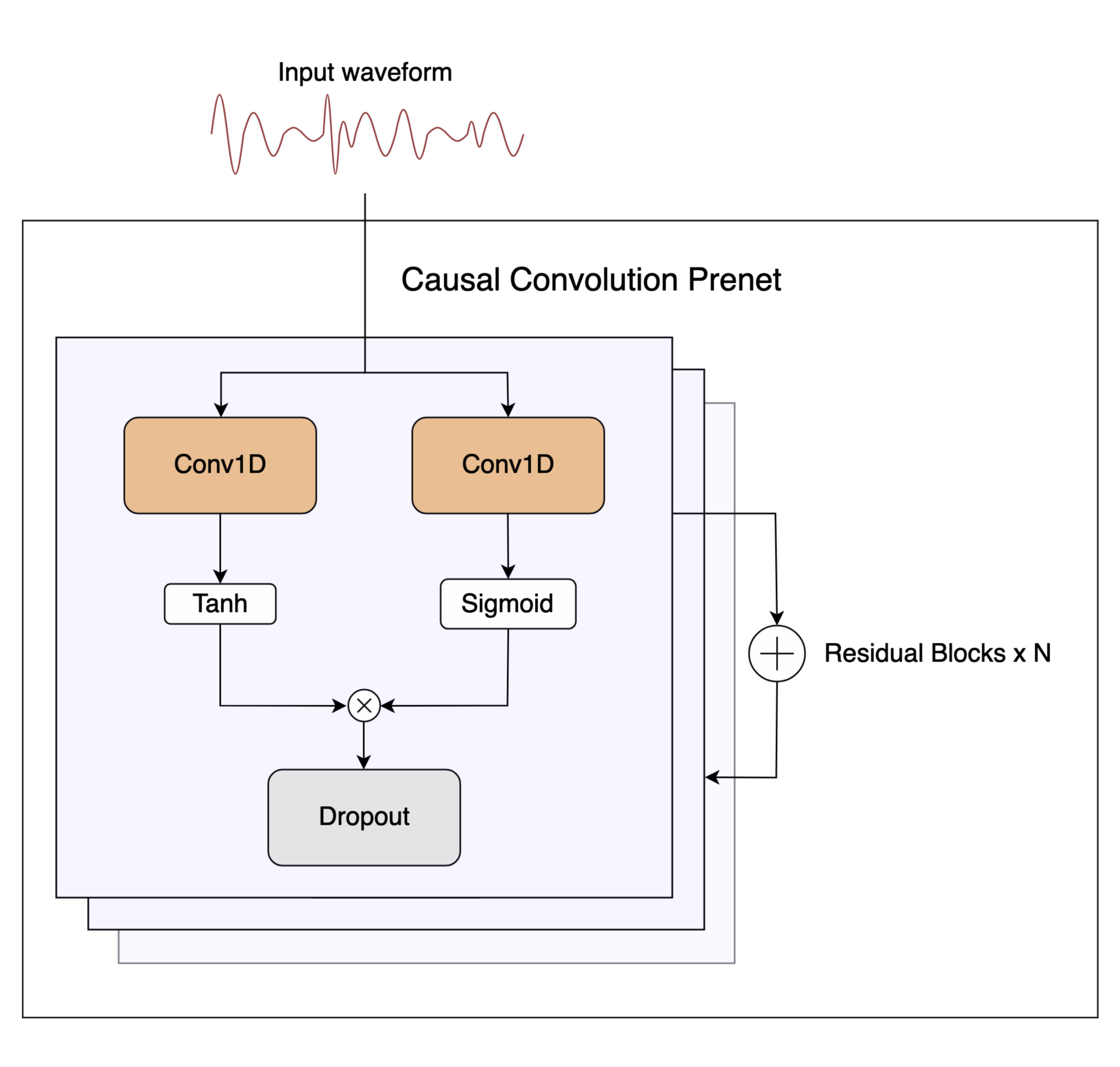}
\caption{Cached Convolution Prenet}
\end{subfigure}
\caption{Generator (figure 1a)and Causal Convolution Prenet architecture (figure 1b). For details on the DCC Encoder and Transformer Decoder architectures, see the Waveformer paper. Note that the Waveformer's Transformer Decoder takes a label query vector as input, which we do not use.}
\end{figure}

\subsubsection{Generator}
Our generator is derived from Waveformer's streaming encoder-decoder model. We adopt Waveformer's 512-dimensional encoder and 256-dimensional decoder as the base for our model, though we decrease encoder depth from 10 to 8 layers and decreasing lookahead to 16 samples for lower inference latency and computation speed. Based on the success of causal U-Nets for speech modeling and enhancement\cite{stoller2019sequnet, Ren2021ACU}, we prefix the model with a prenet composed of causal convolutions.

\subsubsection{Discriminator}
We adopt the multi-period discriminator architecture of VITS\footnote{\url{https://github.com/jaywalnut310/vits}}, with discriminator periods of $[2,3,5,7,11,17,23,37]$ inspired by RVC's\footnote{\url{https://github.com/RVC-Project/Retrieval-based-Voice-Conversion-WebUI}} v2 discriminator.

\subsection{Dataset}
We take the LibriSpeech clean 360 hour train split as input to our model\cite{librispeech}. This dataset consists of audio independently recorded by 922 English speakers with diverse speech characteristics and is thus a reasonable starting point for an any-to-one voice conversion system. We hold back a random sample of 2\% of the files in this dataset from the training set to use for validation. We additionally use the dev-clean split, which contains a disjoint set of speakers from the 360 split in order to validate conversion on unseen input speakers.

We generate parallel utterances in the style of a single target speaker by converting the LibriSpeech files with an RVC v2 model trained on 39 minutes of audio from LibriSpeech speaker 8312, obtained from the librivox.org website. We fine tune a 32k RVC v2 base model\footnote{\url{https://huggingface.co/lj1995/VoiceConversionWebUI/tree/main/pretrained\_v2}} for 325 epochs on the target speaker data, using the RMVPE pitch extraction method \cite{wei2023rmvpe}. The typical RVC pipeline includes a step where encoded input speaker data is mixed with encoded target speaker data retrieved from indexed ground-truth data. We choose to omit this step because we found it to decrease performance and intelligibility without improving conversion quality or resemblance. We downsample the 32kHz converted audio to 16kHz to match the sample rate of the unconverted input.

\subsection{Training}
We trained our model for 500,000 steps (53 epochs) on a single RTX 3090 GPU for 3 days at batch size 9. We used an AdamW optimizer and an exponential learning rate scheduler with gradient normalized to 1 to stabilize training. We set the learning rate to 5e-4, learning rate decay to 0.999, AdamW momentum to 0.8, 0.999, and AdamW epsilon to 1e-9.

\subsubsection{Loss}
Our discriminator uses the same loss as the discriminator from VITS. Our generator uses a weighted sum of the VITS generator and feature loss as well as mel spectrogram and self-supervised speech representation based losses. The mel spectrogram loss is derived from the VITS mel loss, though we replace the VITS implementation with multi-resolution mel spectrogram loss from the auraloss library\cite{steinmetz2020auraloss}. The self-supervised representation loss is inspired by Close, et al. (2023)\cite{Close_2023}, which found that loss based on L1 distance between features encoded by the pretrained fairseq HuBERT Base model was effective for speech enhancement.

\subsection{Inference}
\label{subsec:inference}
The LLVC streaming inference procedure follows Waveformer's chunk-based inference with lookahead. A single chunk is composed of $ dec\_chunk\_len * L$ samples. Inference additionally requires lookahead of $2L$ samples, for a total latency of 
\begin{equation}\label{eq:1}
    (dec\_chunk\_len * L + 2L) / F_s
\end{equation}
seconds, where $F_s$ is the audio sample rate in Hz. It is also possible to run the network with $N$ chunks at a time, increasing latency in order to improve the real-time factor of conversion. The file \texttt{infer.py} in the associated Github repository demonstrates the implementation of streaming inference at variable latency.

\section{Experiments}
In addition to the architecture described above, we trained two additional variants of our model. Hyperparameters for our runs can be found in the .json configuration files in the experiments/ directory of the linked repository.

\subsection{No causal convolution prenet}
The causal convolution prenet adds latency to the model's forward pass, so we performed an ablation run to test its impact on output quality. Hyperparameters and training steps were identical to that of the main LLVC model. We label this experiment LLVC-NC in our comparisons.

\subsection{Hifi-GAN discriminator}
Compared to the VITS discriminator, the HiFi-GAN discriminator uses fewer multi-period sub-discriminators at smaller period sizes, and more multi-scale sub-discriminators at larger period sizes. We reduced the training batch size to 7 but otherwise keep hyperparameters and training step count identical to LLVC. We label this experiment LLVC-HFG in our comparisons.

\section{Results}
\subsection{Evaluation Dataset}
We used the LibriSpeech test-clean files as input for conversion. We used N-second clips from the LibriSpeech dataset for speaker 8312 to test quality and self-similarity of the ground truth dataset. 
\subsection{Comparison}
We select two models for comparison with LLVC with the criterion of minimizing inference latency on CPU.
\begin{itemize}
\item No-F0 RVC: Pitch estimation creates a performance bottleneck for RVC, but the RVC developers provide pre-trained models that do not take pitch as input. We fine-tuned the RVC v2 32k no-f0 models on the 39 minutes of speaker 8312 data for 300 epochs.
\item QuickVC: We fine-tuned the pre-trained QuickVC model linked in the official repository for 100 epochs on the 39 minutes of speaker 8312 data downsampled to 16kHz.
\end{itemize}

\subsection{Performance}
All models were evaluated on a Intel(R) Core(TM) i9-10850K CPU @ 3.60GHz. For No-F0 RVC and QuickVC, we aimed to achieve the lowest latency and highest amount of context that would allow the models to consistently run at above 1x real-time: a new content window of 100ms with a context buffer of 1024 samples for No-F0 RVC, and a window of 50ms and 2048 samples for QuickVC. LLVC was tested with the smallest new content window that the architecture could accommodate: about 15ms, as per \ref{eq:1}.

We obtain performance numbers by averaging inference latency and the real-time factor (RTF) for conversions performed on the 2620 files LibriSpeech test-clean dataset, where RTF is the seconds of speech generated in $1$ second of wall time. LLVC and LLVC-HFG have identical generator architectures, so differences in performance have no bearing on the efficiency of the underlying models. The lowest end-to-end latency and highest RTF scores have been bolded.
\begin{table}[H]
  \centering
  \begin{tabular}{ c|c|c }
    \hline
          & End-to-End Latency (ms) & RTF \\
    \hline
    No-F0 RVC & 189.772 & 1.114    \\
    QuickVC & 97.616 & 1.050      \\
    LLVC (ours)     & 19.696   & 2.769 \\
    LLVC-NC (ours) & \textbf{18.327} & \textbf{3.677} \\
    LLVC-HFG (ours) & 19.563 & 2.850 \\
    \hline
  \end{tabular}
  \label{tab:perf}
\end{table}

\subsection{Naturalness and Target-Speaker Similarity}
We followed Guo et al. (2023)\cite{guo2023quickvc} to obtain Mean Opinion Scores (MOS) for naturalness and similarity to the target speaker of the converted speech. We recruited subjects on Amazon Mechanical Turk. 15 subjects evaluated naturalness of 4 utterances from the dataset and 4 converted utterances per model. 15 subjects individually evaluated the similarity of 2 utterances from the ground-truth dataset and the similarity of 4 converted utterances to 2 clips from the ground-truth dataset. The highest scores among the converted audio are in bold.
\begin{table}[H]
  \centering
  \begin{tabular}{ c|c|c }
    \hline
          & Naturalness & Similarity \\
    \hline
    Ground Truth & 3.7 & 3.88 \\
    No-F0 RVC & 3.58  & 3.35    \\
    QuickVC & 3.28 & 3.26      \\
    LLVC (ours)    & 3.78   & 3.83 \\
    LLVC-NC (ours) & 3.73   & 3.7 \\
    LLVC-HFG (ours) & \textbf{3.88}   & \textbf{3.9} \\
    \hline
  \end{tabular}
  \label{tab:mos}
\end{table}
\subsection{Objective Metrics}
We use the Resemblyze\footnote{\url{https://github.com/resemble-ai/Resemblyzer}} and WVMOS\footnote{\url{https://github.com/AndreevP/wvmos}} libraries\cite{Andreev_2023} in order to obtain metrics for target-speaker similarity and quality for the entire test-clean dataset. We obtain a baseline for comparison by evaluating 10 different 10-second clips from the ground truth against each other. The highest scores among the converted audio have been bolded.
\begin{table}[H]
  \centering
  \begin{tabular}{ c|c|c }
    \hline
          & Resemblyze & WVMOS \\
    \hline
    Ground Truth & 0.898 & 3.854 \\
    No-F0 RVC & \textbf{0.846} & 2.465 \\
    QuickVC\cite{guo2023quickvc}   & 0.828 & 2.828 \\
    LLVC (ours)    & 0.829   & 3.605 \\
    LLVC-NC (ours)   & 0.821  & \textbf{3.677} \\
    LLVC-HFG (ours) & 0.819 & 3.543 \\
    \hline
  \end{tabular}
  \label{tab:objective}
\end{table}

\section{Conclusion and Further Work}
\label{sec:conclusion}
Our work demonstrates the feasibility of ultra-low-latency low-resource voice conversion. LLVC is able to run in a streaming manner on devices that lack a dedicated GPU such as laptops and mobile phones.

We performed dataset preparation and training on a single consumer-grade GPU, using data and pretrained models freely available online. While we trained our own RVC v2 model, any pretrained RVC v2 model can be used to create a dataset for LLVC training. By open-sourcing our code, we hope to provide a broadly accessible option for creating and using real-time voice changing models.

Our choice of training data contained only clean English speech, even though our method of constructing the parallel dataset is language-independent and relatively robust to noise. Incorporating multi-lingual and noisy speech could create a model that generalizes better across diverse speakers. Conversely, our model could be fine-tuned on a dataset comprised of only a single input speaker converted to a target voice in order to create a personalized voice conversion model.

\section*{Acknowledgments}
Koe AI\footnote{\url{https://koe.ai/}} provided compute and funding for this research. We thank Dr. Kyle Wilson at Washington College and Dr. Lorenz Diener for providing feedback on the first draft of the preprint.

\bibliographystyle{plainnat} 
\bibliography{references}  

\end{document}